\documentclass[aps,prl,twocolumn,nopacs,superscriptaddress]{revtex4}
\usepackage{graphicx}
\usepackage{verbatim}
\usepackage{amsmath}
\usepackage{sidecap}
\usepackage{epstopdf}

\begin{document}

\title{Spectroscopic determination of the atomic f-electron symmetry underlying hidden order in URu$_2$Si$_2$}
%\title{Measuring the electronic excitation spectrum of URu$_2$Si$_2$ to understand the atomic basis of hidden order}
%\title{Observation of a moderately correlated atomic state underlying hidden order in URu$_2$Si$_2$}

%\title{Determination of the atomic state basis and energetics that underlie hidden order in URu$_2$Si$_2$}
%\title{Identifying the atomic state basis and energetics that underlie hidden order in URu$_2$Si$_2$}
%\title{Excitations of the correlated electronic wavefunction in the hidden order compound URu$_2$Si$_2$}
%\title{Measuring the correlated electronic state and excitation spectrum on uranium in URu$_2$Si$_2$}
%\title{Observation of correlated but isotropic electronic states on uranium in URu$_2$Si$_2$}
%\title{Experimental determination of atomic degrees of freedom and electronic Hamiltonian terms underlying hidden order in URu$_2$Si$_2$}
%\title{Charge excitation spectrum of a hidden $5f^2$ uranium lattice in URu$_2$Si$_2$}

\author{L. Andrew Wray}
\email{lawray@nyu.edu}
\thanks{Corresponding author}
\affiliation{Department of Physics, New York University, New York, New York 10003, USA}
%\affiliation{NYU-ECNU Joint Physics Research Institute, East China Normal University, Shanghai 200062, China}
\affiliation{Stanford Institute for Materials and Energy Sciences, SLAC National Accelerator Laboratory, 2575 Sand Hill Road, Menlo Park, CA 94025, USA}
\affiliation{Advanced Light Source, Lawrence Berkeley National Laboratory, Berkeley, CA 94720, USA}
\author{Jonathan Denlinger}
\author{Shih-Wen Huang}
\affiliation{Advanced Light Source, Lawrence Berkeley National Laboratory, Berkeley, CA 94720, USA}
\author{Haowei He}
\affiliation{Department of Physics, New York University, New York, New York 10003, USA}
%\author{Jacques Flouquet}
%\affiliation{INAC/SPSMS, CEA-Grenoble, 38054 Grenoble, France}
%\author{Andrew D. Huxley}
%\affiliation{School of Physics, James Clerk Maxwell Building, King's Buildings, Edinburgh, U.K.}
\author{Nicholas P. Butch}
\affiliation{Center for Nanophysics and Advanced Materials, Department of Physics, University of Maryland, College Park, MD 20742, USA}
\affiliation{NIST Center for Neutron Research, National Institute of Standards and Technology, 100 Bureau Drive, Gaithersburg, MD 20899, USA}
\author{M. Brian Maple}
\affiliation{Department of Physics, U. of California, San Diego, La Jolla, CA 92093, USA}
\author{Zahid Hussain}
\author{Yi-De Chuang}
\affiliation{Advanced Light Source, Lawrence Berkeley National Laboratory, Berkeley, CA 94720, USA}

\begin{abstract}

The low temperature hidden order state of URu$_2$Si$_2$ has long been a subject of intense speculation, and is thought to represent an as yet undetermined many-body quantum state not realized by other known materials. Here, X-ray absorption spectroscopy (XAS) and high resolution resonant inelastic X-ray scattering (RIXS) are used to observe electronic excitation spectra of URu$_2$Si$_2$, as a means to identify the degrees of freedom available to constitute the hidden order wavefunction. Excitations are shown to have symmetries that derive from a correlated $5f^2$ atomic multiplet basis that is modified by itinerancy. The features, amplitude and temperature dependence of linear dichroism are in agreement with ground states that closely resemble the doublet $\Gamma_5$ crystal field state of uranium.

\end{abstract}

% \pacs{}

\date{\today}

\maketitle

%\begin{SCfigure*}
%\centering
%\includegraphics[width = 12.5cm]{Figure1}

\begin{figure}
\centering
\includegraphics[width = 8.6cm]{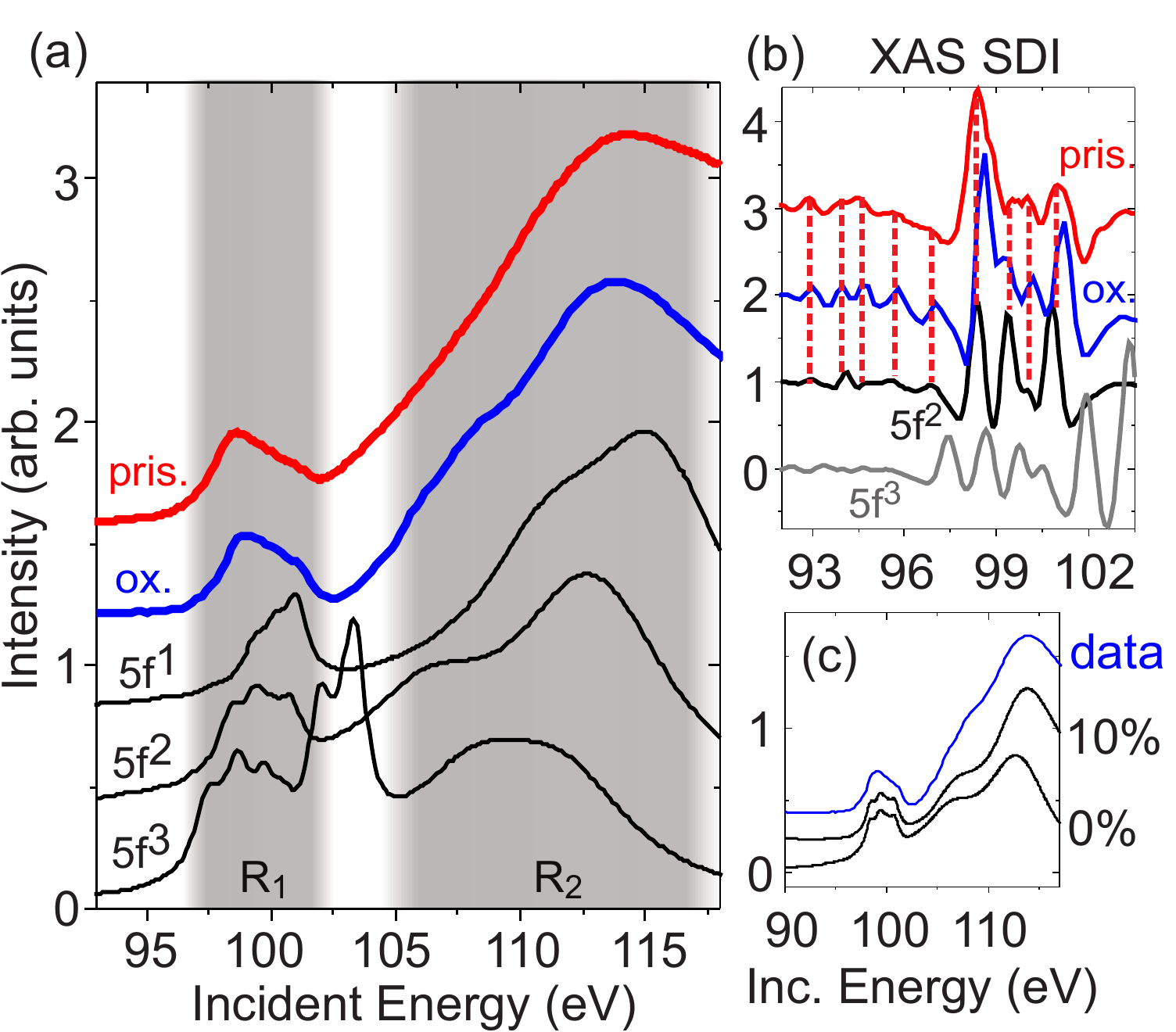}
\caption{{\bf{Resonance on uranium}}: (a) X-ray absorption measured at T=10K for (blue) the oxidized sample and (red) the pristine surface of URu$_2$Si$_2$, with incident polarization $\theta=60^o$ from the [001] crystal face. Black curves show simulations for $5f^1$, $5f^2$ and $5f^3$ entangled states. (b) Negative second derivative images (SDI) of XAS spectra in panel (a). Drop-lines trace a one-to-one correspondence between local maxima in the experimental data and the $5f^2$ simulation. Curves are scaled for feature visibility and vertically offset by integer units. (c) Applying a weak (10$\%$) Fano effect to the calculated $5f^2$ spectrum improves correspondence with XAS data.}
\label{fig:UraniumF2}
\end{figure}

%\begin{figure}[t]
%\includegraphics[width = 8.65cm]{Figure1}
%\caption{{\bf{Resonance on uranium}}: (a) Different classes of electronic states that may be realized within URS are illustrated.  Low energy states are illustrated at left, and characteristic excitations are shown at right. (b) X-ray absorption measured at T=10K for (blue) the oxidized sample and (red) the pristine surface of URS. Black curves show simulations for $5f^1$, $5f^2$ and $5f^3$ entangled states. (inset) Applying a weak (10$\%$) Fano effect to the calculated $5f^2$ spectrum improves correspondence with XAS data. (c) The negative second derivative image (SDI) of XAS spectra in panel (b) shows a one-to-one correspondence between experimental data and features in the $5f^2$ simulation.}
%\label{fig:UraniumF2}
%\end{figure}

The low temperature hidden order (HO) phase of URu$_2$Si$_2$ has been a mystery for more than 25 years, and is widely anticipated to represent a novel many-body quantum state. When cooling through T$_{HO}$=17.5K the material undergoes a second order phase transition, with a large loss of entropy that cannot be immediately explained by observed changes in the electronic structure \cite{entropyChange,entropyUnexplained,AmitsukaRXS,WalkerRXS}. Pinpointing the microscopic cause of this entropy change is challenging because basic properties of the atomic scale wavefunction are not decisively known. Experiments differ on whether the uranium valence state is closer to $U^{4+}$ ($5f^2$) \cite{inelastNeutron1,inelastNeutron2} or $U^{3+}$ ($5f^3$) \cite{EELS2010}. Proposed models have considered a wide range of local \cite{firstCEFmodel,Gamma5doublet1,Gamma5doublet2,Gamma5doublet3,Gamma3early,local2,local3,Gamma3review,KotliarNatPhys,FlouquetSimpleAnswer,local6,HastaticNature,HastaticReview} and itinerant \cite{nonlocal1,nonlocal2,mydoshBigTheoryPRB,nonlocal3,nonlocal3p5,nonlocal4,nonlocal5,nematicRank5,DasSOdensityWave} low energy state bases for $5f$ electrons, and explored many exciting possibilities for the ``hidden" quantum state. Here, high resolution ($\delta E$$\sim$35meV) resonant inelastic X-ray scattering (RIXS) and X-ray absorption spectroscopy (XAS) are used to measure fundamental excitations created by resonance with the uranium 5d core level (O-edge), to identify what electronic degrees of freedom are relevant for effective models of hidden order, and what degrees of freedom are energetically gapped out.

Measurements were performed on both a pristine crystalline surface cleaved in ultra high vacuum, and a cleaved surface that was oxidized by exposure to air at room temperature, promoting U$^{4+}$ valence. The dominant spectral features observed from \emph{both} surfaces are shown to derive from the excitations of a freestanding $5f^2$ $U^{4+}$ atom, revealing that atomically correlated Hund's rule interactions play a key role in determining the electronic degrees of freedom that can contribute to the hidden order state. However, some low energy excitations of the pristine surface are found to be extremely short lived, implying that the symmetries they represent are not strictly eliminated from the hidden order ground state. Linear dichroism in the XAS spectrum is consistent with the crystalline electric field (CEF) \emph{doublet} state $\Gamma_5$, but inconsistent with CEF \emph{singlet} ground states that have been predicted as the basis of hidden order (e.g. $\Gamma_1$, $\Gamma_2$, $\Gamma_3$).

\begin{figure}[t]
\includegraphics[width = 8.65cm]{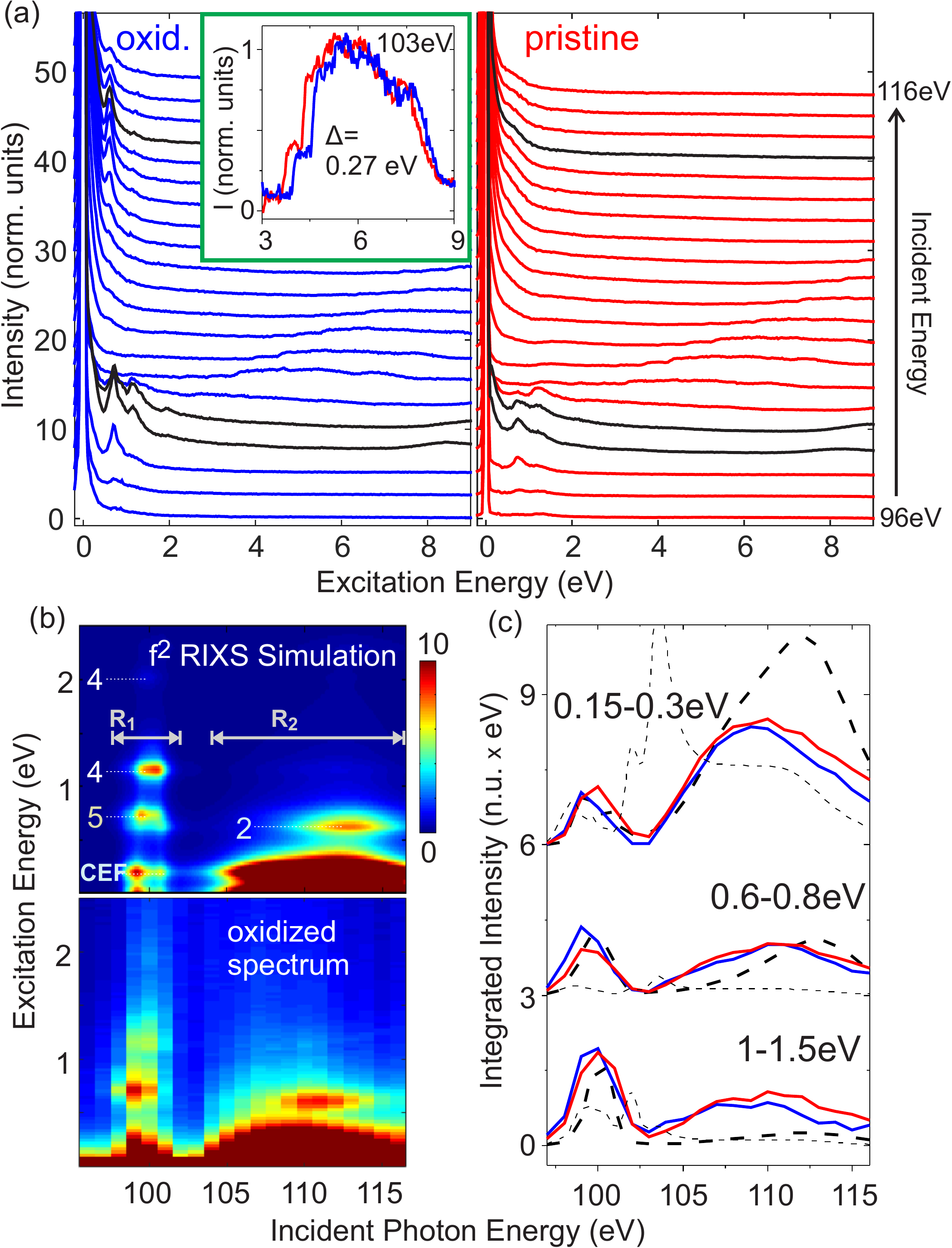}
\caption{{\bf{RIXS spectra of URu$_2$Si$_2$}}: (a) RIXS spectra for the (blue) oxidized and (red) pristine samples are measured with incident photons from (bottom curve) 96eV to (top curve) 116eV, with a 1eV step. Representative curves at the $R_1$ ($h\nu$=99, 100eV) and $R_2$ (h$\nu$=113 eV) resonances are traced in black, and an inset shows the emission line feature as measured at 103eV. (b) The air exposed RIXS spectra are compared with a $5f^2$ simulation, which is labeled by the total angular momentum (J) quantum numbers of excited states. (c) Integrated scattering intensity in the 0.15-0.3, 0.6-0.8, and 1.0-1.5eV excitation energy windows is vertically displaced and plotted as a function of incident energy for (red) pristine URu$_2$Si$_2$ and (blue) the oxidized surface. Intensity from the pristine surface is multiplied by a factor of 1.5. Dashed lines show simulations for (thick) $5f^2$ and (thin) $5f^3$.}
\label{fig:RIXSprof}
\end{figure}

Measuring XAS at the O-edge reveals a fingerprint of how the valence electronic structure projects onto coherent core hole states \cite{deGrootInvisible,5fdichroismME}. Colored curves in Fig. \ref{fig:UraniumF2}(a) show XAS measured on the pristine and air exposed samples within the hidden order state (T=10K). Intensity is divided between two principle regions labeled $R_1$ (97-102eV) and $R_2$ ($\sim$104-118eV), split mostly by $5d-5f$ spin exchange interactions. Comparing the XAS measurements with black curves representing atomic multiplet (AM) calculations for isolated $5f^1$, $5f^2$, and $5f^3$ uranium atoms shows reasonable correspondence with the $5f^2$ multiplet structure, but other than the $R_1$ and $R_2$ features themselves, there are no prominent well resolved peaks that would provide a basis for detailed comparison. The $5f^3$ simulation is relatively incompatible, because its intensity maximum at $\sim$103eV coincides with a local minimum of the experimental curve. Most analysis of XAS in this paper will focus on the $R_1$ region, because the Fano effect identified in Ref. \cite{AugerFano} significantly influences total electron yield (TEY) XAS measurements on shorter-lived resonance states in the $R_2$ region (see Fig. \ref{fig:UraniumF2}(c)).

The technique of second derivative imaging (SDI) is used in Fig. \ref{fig:UraniumF2}(b) to more sharply resolve component features within the XAS spectra. The SDI spectrum of the air exposed sample reveals nine features near $R_1$ (see blue curve), including highly reproducible \cite{SM} fine structure below the $R_1$ energy region. These peaks have a one to one correspondence with features in the $5f^2$ calculation, as might be expected for an oxidized surface with U$^{4+}$ nominal valence. More remarkably, features of the pristine sample also have a one-to-one correspondence with SDI local maxima seen from the air exposed sample (see red drop-lines), although the feature at 98.5 eV appears to be split into two components. SDI features of the pristine surface are broader and slightly lower in energy than those of the air-cleaved surface.

%Feature intensities are not identical, suggesting some variation in the occupation of $5f^2$ symmetries (see Fig. 4 discussion).

% that the pristine and oxidized surfaces ground states incorporate different detailed symmetries, which are not a perfect match for the ground state of the CEF multiplet. SDI features of the pristine surface are broader and slightly lower in energy than those from the air-cleaved surface.

%No features from the $5f^3$ calculation (grey curve) can be definitively identified. Slightly different feature intensities between the air exposed and pristine samples suggest that their ground states incorporate different $5f^2$ symmetries, as will be corroborated by other data in Fig. 3-4.

The correspondence of XAS with a $5f^2$ multiplet strongly suggests that the atomic scale electronic degrees of freedom available for incorporation in the hidden order ground state come largely from the low energy states of a $5f^2$ manifold. However, it is also possible that $5f^3$ states may be underrepresented in the XAS spectrum if they are relatively itinerant and persist on the core hole site for too short a time to yield sharp excitation features. Technical concerns for relating O-edge XAS to valence estimates from other techniques are discussed in the online Supplemental Material (SM) \cite{SM}.

%, due to the combined itinerancy of three electrons rather than two

\begin{figure}
\centering
\includegraphics[width = 8.7cm]{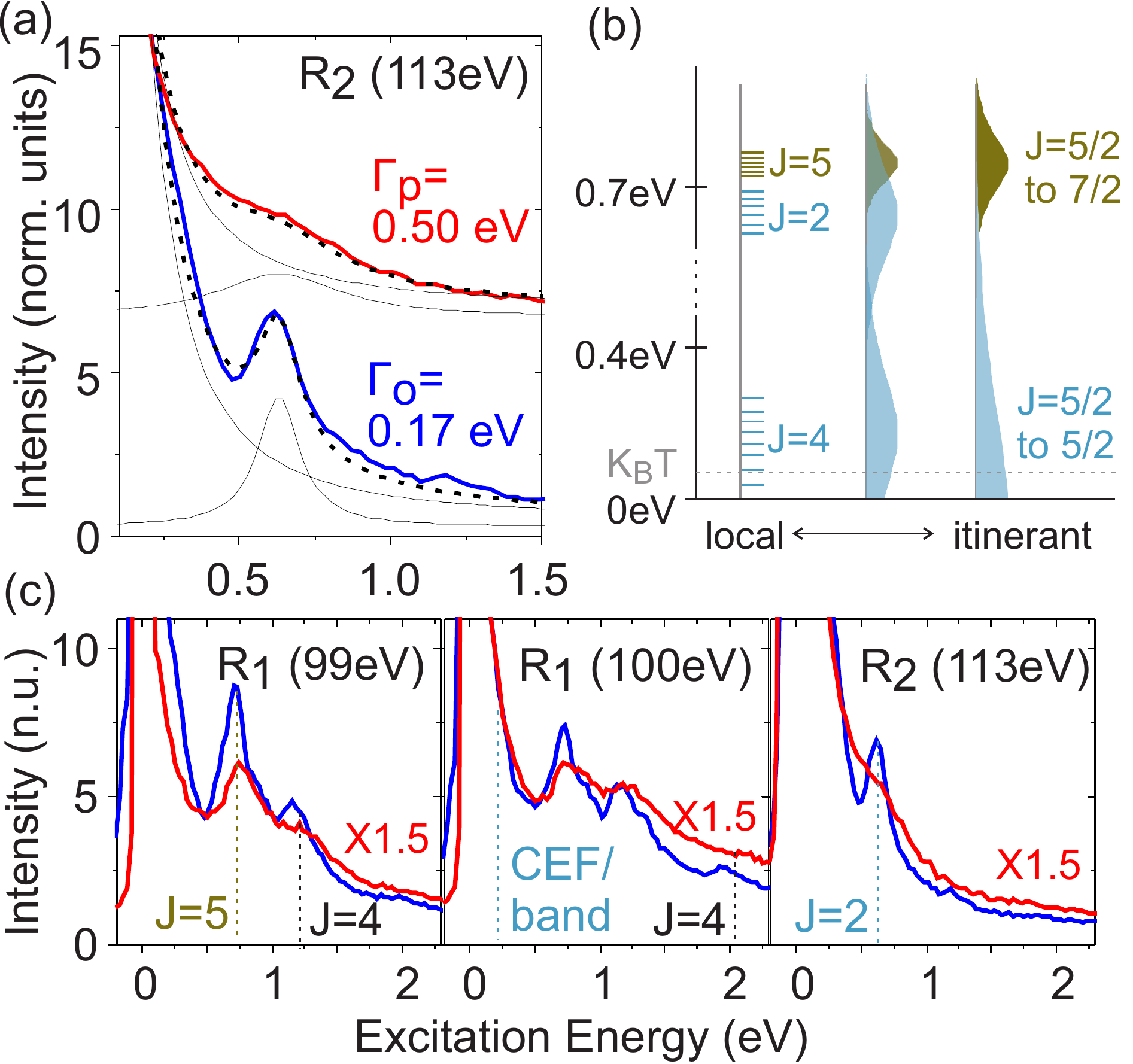}
\caption{{\bf{Excitations and excited state symmetries}}: (a) The $R_2$ J=2 excitation measured from the (red) pristine and (blue) oxidized surfaces is normalized to the Si L-edge fluorescence and fitted with a single Lorentzian on a curved background. The Lorentzians used for each sample have identical area. (b) Schematic of excited states accessible from single atom transitions in a (left) local, (middle) semi-itinerant, and (right) highly itinerant pictures. (c) Representative RIXS curves and calculated excitation symmetries are shown for excitations at $R_1$ and $R_2$. Intensity is normalized to Si L-edge emission as in panel (a), and has been multiplied by a factor of 1.5 following normalization for the pristine surface.}
\label{fig:RIXSanalysis}
\end{figure}

Performing RIXS at the same incident photon energies reveals low energy excitations that are left behind when the O-edge core hole decays, and provides much greater bulk sensitivity than XAS \cite{SM}. The RIXS scattering process involves the short lived excitation of a core electron to the valence level, followed by another electronic transition that fills the core hole and leaves behind a low energy excitation \cite{ButorinReview,KotaniRIXSreview,AmentRIXSReview,WrayFrontiers}. Depending on how correlated or itinerant the electronic system is, electronic excitations on uranium may resemble band transitions between itinerant single electron states, which have J=5/2 or 7/2 angular momentum, or may be dominated by excitations that change the coherent relationship between locally entangled electrons (see diagram in Fig. \ref{fig:RIXSanalysis}(b)). Regardless of the entangled or itinerant nature of low energy electrons, the angular momentum coupling between a $5d$ core hole and the $5f$ valence electrons has an extremely large $\sim$15eV energy scale, leading to coherent resonance states with strong entanglement between valence electrons and the core hole, and qualitatively different spectra from previous studies of $3d$ and $4d$ core hole resonance \cite{EELS2010,M5M4,SM}.
%***Add my Frontiers paper after AmentRIXSReview!

%The RIXS emission line features discussed further below also imply that the number of charge carriers with predominantly 5f-symmetry is small.

The incident energy dependence of RIXS spectra for the pristine and air exposed samples is shown in Fig. \ref{fig:RIXSprof}(a), with scattering intensity plotted in normalized units of the Si L-edge fluorescence amplitude. Both profiles show the same qualitative features, but there are numerous quantitative differences. In each case, almost all intensity at $E>3 eV$ energy loss is found in a broad, linearly dispersive feature termed an emission line, which resonates at incident photon energies $h\nu \gtrsim 102eV$ \cite{SM}. Corresponding features of the emission line are shifted by $\Delta=270 meV$ between the pristine and air exposed samples (see Fig. \ref{fig:RIXSprof}(a) inset). Emission lines disperse with a slope of 1, and are thought to approximately represent the density of single particle electronic states below the Fermi level that can transition to fill the core hole \cite{JinghuaCKalpha,JinghuaReview,HancockRIXSmetal}. In this picture, oxidation lowers the energy of ligand band structure (shifting features to larger energy loss) by reducing the surrounding charge density. No feature at the leading edge of the emission line is seen to be lost as a consequence of oxidation, suggesting that oxidation has little effect on the occupied electronic symmetries.
%spread over 5eV
%***also cite Japanese group from gunma university

The relatively sharp air exposed RIXS spectrum is compared with a $5f^2$ AM model in Fig. \ref{fig:RIXSprof}(b), and again establishes a one to one correspondence of features, with details very similar to the RIXS spectrum of uranium oxide (UO$_2$) \cite{ButorinReview}. The total angular momentum (J) quantum number of atomic multiplet excitations is labeled on the simulation. The CEF excitations will manifest as band excitations in an itinerant picture, and have therefore been set to 0.2 eV to correspond with the approximate energy scale expected for low energy uranium band transitions (between J=5/2 bands). The prominent 0.6 eV excitation visible at $R_2$ has J=2 angular momentum, and is unrelated to the nearly degenerate J=5 feature seen at $R_1$. Comparing the incident energy dependence of the three lowest energy features shows that each can be qualitatively explained by a $5f^2$ atomic multiplet model (thick dashed line in Fig. \ref{fig:RIXSprof}(c)), and has no distinct correspondence with a $5f^3$ model (thin dashed line).

Excitations from 0.15 to 0.3 eV can account for most of the resonant intensity that the model attributes to CEF $ff$ excitations (see Fig. \ref{fig:RIXSprof}(c), top curve). These CEF transitions are created by changing the coherent relationship between two electrons in J=5/2 single particle states on the same atom, and in an itinerant picture the electrons will rapidly delocalize into excitations between the J=5/2 bands that intersect the Fermi level. RIXS curves show no sharp features in this energy range (Fig. \ref{fig:RIXSanalysis}(c)), and neutron scattering does not reveal CEF excitations \cite{ButchNeutron}, so one may conclude that CEF transitions are observed in the measurement, but are too short lived to be considered stable collective modes. The intensity at $E<0.3eV$ may be best interpreted as a combination of coherent band transitions and more complex dynamical modes triggered by a local change in the orientation of spin-orbit coupled angular momentum.

Other RIXS excitations of the pristine sample generally appear to be 5-10$\%$ higher in energy than those of the air exposed sample. A fitting of the atomic energetics (spin orbit coupling and 2-particle Coulomb interactions) is presented in the SM, identifying very standard degrees of renormalization (5-20$\%$) relative to first principles Hartree-Fock values. Although these parameters are slightly optimized to match experimental excitation energies, qualitative features of the calculated XAS and RIXS spectra are essentially parameter independent.

In distinguishing between itinerant and local pictures of uranium physics, it is useful to focus on the J=5 and J=2 excitations found close to 0.7eV. In an itinerant picture, the J=5 mode appears as a gapped inter-J (J=5/2 to 7/2) band excitation, while the J=2 excitation melts into the gapless J=5/2 to J=5/2 intra-J excitation continuum (see Fig. \ref{fig:RIXSanalysis}(b)). Both of these excitations are seen clearly in the air exposed sample, but the J=2 feature seen from the pristine sample is roughly three times broader (see Fig. \ref{fig:RIXSanalysis}(a)), with a fitted inverse lifetime of $\Gamma_v$$\sim$0.5 eV that is comparable to the mode energy. The J=5 excitation has a similarly sharp line shape for each sample (Fig. \ref{fig:RIXSanalysis}(c), left panels), consistent with the picture that the J=5 mode is viable in an itinerant picture, but the J=2 mode decays as rapidly as electrons can delocalize away from the scattering site. The existence of a mostly-gapped J=2 mode in our spectra from URu$_2$Si$_2$ implies that electrons reside long enough on the same atom to enter coherent states composed primarily (but not exclusively) of the ground state manifold obtained from Hund's rule energetics for $5f^2$ uranium, which has the angular momentum quantum number J=4.

% However, it maintains the same integrated intensity, when data are normalized to the Si emission line.

\begin{figure}[t]
\includegraphics[width = 8.65cm]{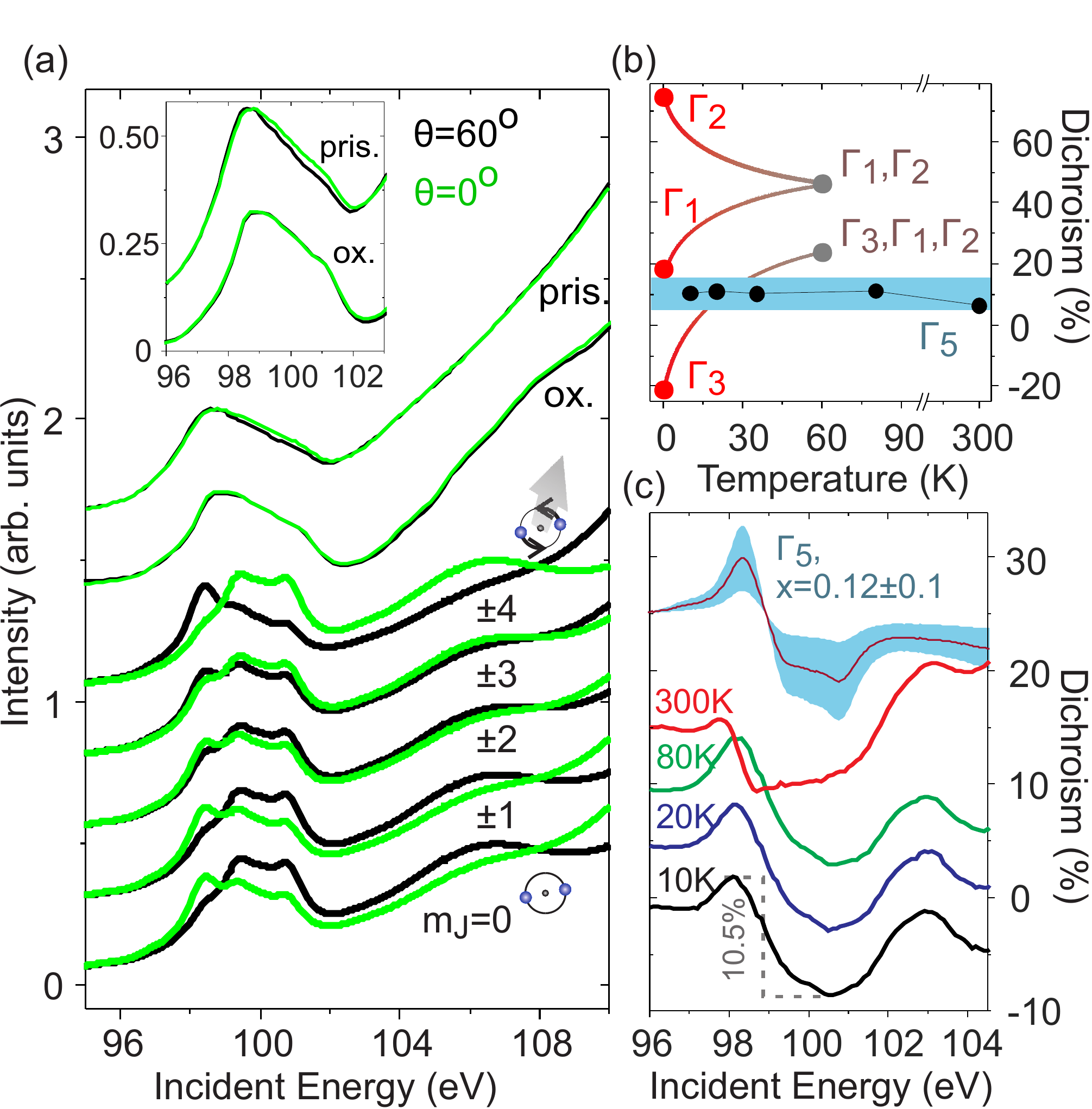}
\caption{{\bf{Dichroism and the atomic polarizability of uranium:}} (a) X-ray absorption in the hidden order state (T=10K) is measured for polarization (green) $\theta=0^\circ$ and (black) $60^\circ$ from the [001] crystal face, and compared with (thick curves) $5f^2$ simulations for pure $m_J$ eigenstates. A blow-up of the experimental curves is shown in the inset. (b) The temperature dependence of dichroic contrast is plotted, and a turquoise shaded region indicates values allowed for the  $\Gamma_5$ state with a mixing parameter of x=0.12$\pm$0.1. Curved lines trace the temperature dependence expected for other proposed CEF wavefunctions of uranium. (c) XAS dichroism curves ($I(60^\circ)-I(0^\circ)$) are shown in percentage units of $R_1$ intensity\cite{SM} and compared with (top) a simulation for the  $\Gamma_5$ state. Curves are offset by integer multiples of (a) 0.25, (a, inset) 0.083, and (c) 5$\%$.}
\label{fig:HamAndPsi}
\end{figure}

Provided that this is the case, an immediate next question is what superposition of J=4 moment states ($m_J$ states) is occupied, and if this superposition changes as a function of temperature. A sizable fraction of the URu$_2$Si$_2$ literature has invoked CEF energy levels, which are coherent $m_J$ superpositions, as a key component in the low energy framework for understanding hidden order and the temperature dependence of susceptibility above hidden order. In scattering data, the occupation of $m_J$ states for $5f^2$ systems can be directly evaluated from linear dichroism in O-edge X-ray absorption spectroscopy \cite{5fdichroismME}, as seen from the simulated XAS curves in Fig. \ref{fig:HamAndPsi}(a) (thick curves). Dichroic contrast flips sign for wavefunction components with $|m_J|\geq3$ and $|m_J|\leq2$, providing a qualitative metric to differentiate between models with a $|m_J|$=2 ground state \cite{Gamma3early,Gamma3review} and ground states dominated by $|m_J|$=3 or 4 components \cite{local3,firstCEFmodel,Gamma3review,Gamma5doublet1,Gamma5doublet2,Gamma5doublet3,KotliarNatPhys}.

Measuring X-ray absorption from the pristine sample with contrasting polarization conditions yields spectra that differ by $\sim$5$\%$ of the maximum $R_1$ intensity (Fig. \ref{fig:HamAndPsi}(c)). Spectra from the air exposed sample show no reproducible dichroism other than a slight difference in the intensity tail from $R_2$, and will not be discussed in further detail. The dichroic difference spectrum of the pristine sample has a fluctuating pattern over $R_1$, with two peaks near 98 eV and 102-103 eV, and a prominent valley from 99-101 eV. All of these features are universally present in simulations with ground state symmetry dominated by a $|m_J|>2$ basis (for example, the top curve in Fig. \ref{fig:HamAndPsi}(c)), and appear with opposite sign in models with a ground state dominated by $|m_J|\leq2$. Raising temperature yields no significant change across the T=17.5K hidden order transition, and the peak-to-valley difference in the dichroic spectrum keeps roughly the same amplitude ($\sim$11$\%$) up to 80K, as plotted in Fig. \ref{fig:HamAndPsi}(b). At room temperature, the dichroic spectrum is significantly distorted and has reduced amplitude, but retains qualitatively similar features. 

A J=4 multiplet in the tetragonal crystal structure of URu$_2$Si$_2$ has 5 singlet states, of which three have been proposed by different studies as the hidden order ground state. These states are termed the $\Gamma_1$ ($|m_J|$=0,4 components), $\Gamma_2$ ($|m_J|$=4) and $\Gamma_3$ ($|m_J|$=2), and a gap of roughly $E\sim6 meV = 2\times K_B T_{HO}$ is required between the ground state and the next excited state to explain the sharp rise in magnetic susceptibility as temperature is raised across the hidden order transition \cite{entropyChange}. Temperature dependence of dichroism as these states are thermalized into their magnetically polarizable manifolds is outlined in Fig. \ref{fig:HamAndPsi}(b), using the $\Gamma_1$ mixing angle obtained numerically in Ref. \cite{KotliarNatPhys}. The lack of temperature dependence in our data (black points in Fig. \ref{fig:HamAndPsi}(b)) is qualitatively incompatible with these CEF hidden order pictures.

A last class of crystal field based models assigns the ground state as a doublet with $\Gamma_5$ symmetry ($|m_J|$=3,1 components). The $\Gamma_5$ atomic ground state is degenerate and magnetically polarizable, and therefore does not directly explain hidden order, but provides a basis for pictures that incorporate interatomic dynamics such as the recent proposal of exotic hastatic order \cite{HastaticNature,HastaticReview}. The XAS dichroism of a $\Gamma_5$ ground state basis depends on the mixing parameter x=$\langle\Psi|(n_{1})|\Psi\rangle$, where $n_{1}$ is the number operator for $|m_J|$$=$$1$ 2-electron configurations. A value of x=0.12$\pm$0.1 yields the simulation plotted at the top of \ref{fig:HamAndPsi}(c), and is close to the value of x=0.2 considered in Ref. \cite{Gamma5doublet1}. The range of dichroic amplitude modeled for $\Gamma_5$ with x=0.12$\pm$0.1 is shaded in turquoise in Fig. \ref{fig:HamAndPsi}(c), and encompasses the data points from pristine URu$_2$Si$_2$. Because the $\Gamma_5$ basis is magnetically polarizable, the suppression of magnetic susceptibility near the hidden order transition must be attributed to interatomic many-body effects, and one does not expect low temperature thermal activation of other CEF states.

In conclusion, these measurements show sharp excitation spectra of URu$_2$Si$_2$, which are definitively linked to the excitation modes of $5f^2$ uranium through comparison with spectra from an oxidized surface. Crystal field excitation intensity, which has been elusive in inelastic neutron scattering, is finally observed, but found not to represent long-lived stable collective excitations. The large energy loss width of crystal field scattering intensity suggests that electrons are sufficiently itinerant to make the simplest rendering of CEF models untenable. However, itinerancy is not quite sufficient to dominate over the energy gap between the J=4 and J=2 atomic states that together give an almost complete basis for $5f^2$ superpositions of the J=5/2 valence band electrons. The preserved energy gap between these modes implies that one should look for the principle components of hidden order in the $m_J$ basis of the J=4 atomic ground state. Linear dichroism is observed in XAS spectra, and analyzed to rule out CEF models that would directly explain hidden order via a singlet uranium ground state. The features, amplitude and temperature of XAS dichroism are found to be highly consistent with a $\Gamma_5$ doublet state that has formed the basis for intriguing many-body models and is compatible with experimental features such as nematicity \cite{okazakiNematic,HastaticReview} and large z-axis susceptibility. More generally, a non-singlet atomic ground state would enable Kondo-lattice physics within the hidden order phase, consistent with the observation of Kondo resonance features by point contact measurements \cite{KondoLatticeNature,Kondo30K}.

\textbf{Acknowledgements:} We are grateful for discussions with K. Wohlfeld. The Advanced Light Source is supported by the Director, Office of Science, Office of Basic Energy Sciences, of the U.S. Department of Energy under Contract No. DE-AC02-05CH11231. Crystal growth and characterization at UCSD was supported by the U.S. DOE under Grant No. DE-FG02-04-ER46105.

%Multipole order is expected \cite{FlouquetSimpleAnswer,nematicRank5}

%<4| has 1.9% pDOS from J=7/2 states

\end{document}